\documentclass[12pt,letterpaper]{article}

\usepackage{amssymb,amsmath,latexsym}  
\usepackage{hyperref}
\usepackage{graphicx}
\usepackage[utf8]{inputenc}
\usepackage[affil-it]{authblk}
\usepackage[english]{babel}

\usepackage{color}
\usepackage{ulem}

\begin{document}

	\title{Acetylene-based frequency stabilization of a laser system for potassium laser cooling.} 
	
	\author[1]{Charbel Cherfan}
	\author[1]{Isam Manai}
	\author[1]{Samir Zemmouri}
	\author[1]{Jean-Claude Garreau}
	\author[1]{Jean-Fran\c cois Cl\'ement}
	\author[1]{Pascal Szriftgiser}
	\author[1]{Radu Chicireanu}
	
	\affil[1]{Universit\'e de Lille, CNRS, UMR 8523 -- PhLAM -- Laboratoire	de Physique des Lasers Atomes et Mol\'ecules, F-59000 Lille, France}

	\date{\today}
	\maketitle

\begin{abstract}
	We demonstrate a laser frequency stabilization technique for laser cooling of potassium atoms, based on saturated absorption spectroscopy in the C-Band optical telecommunication window, using ro-vibrational transitions of the acetylene molecule ($^{12}\rm C_{2}H_{2}$). We identified and characterized several molecular lines, which allow to address each of the potassium D2 (767 nm) and D1 (770 nm) cooling transitions, thanks to a high-power second harmonic generation (SHG) stage. We successfully used this laser system to cool the $^{41}$K isotope of potassium in a 2D-3D Magneto-Optical Traps setup.
\end{abstract}

\section{Introduction}
For laser cooling and trapping experiments, frequency stabilization of the laser system must be ensured, with a stability typically less than the natural linewidth of the transition (a few MHz for alkali atoms). In most cases, this is achieved via sub-Doppler saturated absorption spectroscopy, usually employing the same atomic transition used for laser cooling~\cite{BrunerApplOpt1998, MudarikwaJPhysB2012}. Semiconductor laser diodes in external cavities~\cite{Arnold98,Ricci95},  eventually further amplified with slave diodes or semiconductor tapered amplifiers~\cite{Voigt2001, Nyman06, Stern_10}, are the most common laser sources for laser cooling alkali atoms. Even though those laser setups have been widely implemented over the last three decades in cold atoms experiments, several drawbacks still remain: limited lifetime of tapered amplifiers or poor quality of spatial mode. This last feature implies significant losses of optical power which is crucial for standard experiments of ultracold and quantum degenerate gases. In some cases, powerful lasers are either too costly (as is the case with Ti:Saphire lasers) or unavailable at the desired wavelengths, and frequency conversion (i.e. second harmonic generation or frequency summation) is necessary~\cite{Stern_10, DibouneOptExpress2017}. This solution allows to take advantage of the optoelectronic devices developed for the telecommunications industry in the 1530-1565 nm band. Powerful fiber lasers/amplifiers are available and can be frequency-doubled to the near-infrared (NIR). For laser cooling experiments, two atomic alkali species are suitable for the use of telecom-domain fiber amplifiers: rubidium and potassium. Solutions relying on telecom technologies and second harmonic generation (SHG) have been successfully tested in the case of rubidium, starting from a laser source at 1560 nm~\cite{Thompson:03, Lienhart2007, CarrazApplPhysB2009}.

For potassium, a similar technique is implemented in our setup using a diode laser in the telecom domain, followed by amplification and SHG in a periodically-poled lithium niobate (PPLN) crystal. This allows us to obtain a high laser power at the desired wavelength, close to the D2 cooling transitions at $766.701$~nm~\cite{FalkePRA2006}. On the other hand, ultracold atom laser systems usually require complex amplitude and frequency light control sequences~\cite{Salomon_2013}. To achieve such versatility, two SHG systems are usually developed: a powerful one is dedicated to the atomic laser cooling, and allows dynamic control of the cooling parameters, while a weaker, stationary one is devoted to the frequency locking only~\cite{DibouneOptExpress2017}. However, this solution is cumbersome and expensive. This is why we chose, in our system, to generate simultaneously the cooling and repumper frequencies using a single telecom amplifier. Thus, all cooling frequencies are readily present in a single laser beam, at the output of the SHG stage, eliminating subsequent power losses. Furthermore, the cooling and repumper detunings, as well as their respective power ratio can be varied dynamically at will during the different laser cooling phases~\cite{Charbel2019InPrep}. This feature, however, also makes locking onto potassium saturated absorption, using the SHG output light, quite impractical.

\begin{figure}[!h]
	\centering
	\includegraphics[width=0.8\linewidth]{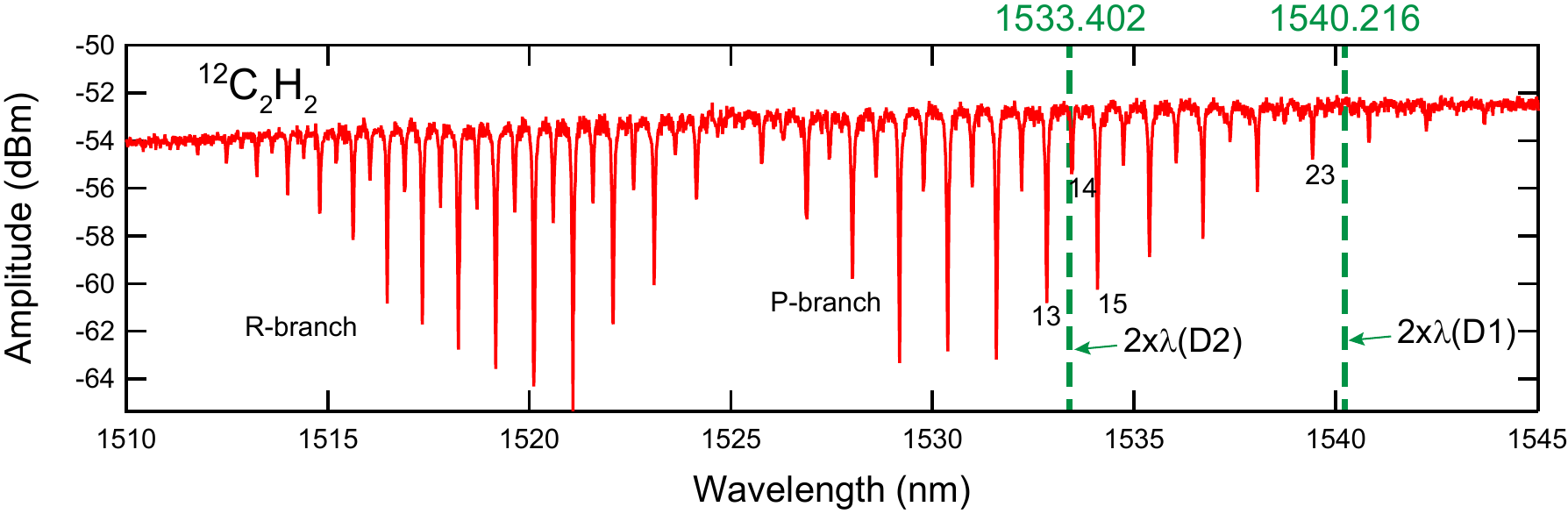}
	\caption{ Spectrum of the Acetylene  $^{12} \rm C_{2}H_{2}$ in the region of $\nu_{1}+\nu_{3}$ band showing the P and R branches. This spectrum is measured experimentally by an Optical Spectrum Analyzer (OSA) containing an acetylene cell, with a resolution of $0.02$~nm and an accuracy below $0.1$~nm (estimated by comparison with~\cite{Nakagawa-96,Acetylene_2001}). The dashed lines (green) correspond to the double of wavelengths of the D1 and D2 transitions of potassium.}
	\label{Acetylene_spectrum}
\end{figure}

To fully preserve the simplicity and flexibility of the laser system, we chose a different approach to the frequency stabilization: using a telecom-based locking scheme. In our case, we chose to use ro-vibrational  transitions of the acetylene molecule, which are close to twice the wavelength of the potassium D2 cooling transitions at $766.701$~nm\cite{FalkePRA2006}. Thus, the frequency stabilization can be completely decoupled from the power amplification and SHG stages. The performance of the system has been proven, by demonstrating the magneto-optical trap (MOT) of potassium atoms, in a 2D+3D MOT system, creating favorable conditions for upcoming cooling stages, paving the way for efficient and reliable potassium Bose-Einstein condensation experiments. In addition, we also tested that our laser stabilization technique can be used to spectroscopically address the D1 transition at 770.108~nm, which can have a particular interest for implementing the  the so-called `gray molasses' technique~\cite{Salomon_2013} and achieving sub-Doppler cooling of potassium.

\section{Experimental setup and results}

\subsection{Acetylene transitions for potassium laser cooling}
The frequency stabilization in our system is realized by using  molecular transitions of the $(\nu_{1}+\nu_{3})$ ro-vibrational band of the $^{12}\rm C_{2}H_{2}$ molecule, the most naturally abundant ($97.7599\%$) isotopologue of acetylene. This molecule contains approximately 50 strong absorption lines in the telecom spectral region, from 1510 nm to 1542 nm~\cite{Nakagawa-96, Acetylene_2001}, as shown in Fig.~\ref{Acetylene_spectrum}. The lines form two distinct branches, P and R, and are labeled by an integer quantum number $n$. The typical frequency interval between consecutive lines ranges from $\simeq 50$, up to $90$~GHz. In $^{12}\rm C_{2}H_{2}$, the intensity of the odd lines is stronger, by typically a factor of three, than the even ones~\cite{Acetylene_2001} (because selection rules related to spin degeneracy and symmetries of the $^{12}$C$_2$H$_2$ molecule).

To determine the transitions of interest for potassium laser cooling, the absorption spectrum of acetylene was recorded using an Optical Spectrum Analyzer (OSA Ando AQ6317B) which contains a reference acetylene cell. As we can see in the Fig.~\ref{Acetylene_spectrum}, the double of the wavelengths of the potassium D lines (vertical dashed lines) are close to different transitions in the P branch. For the potassium D2 line ($766.701$~nm), the closest transition of interest in $^{12}\rm C_{2}H_{2}$ is the $\rm P(14)$, ($1533.461$~nm), with a frequency difference $f_{P(14)}-f_{D2}/2\simeq -7.64$~GHz. However, the $\rm P(14)$ $^{12}\rm C_{2}H_{2}$ line strength is smaller than that of the neighboring $\rm P(13)$ and $\rm P(15)$ lines. For this reason, the utilization of the P(15) line was decided prior to the construction of the setup. In this case, the frequency differences with respect to (half) the potassium D2 frequency are significantly larger: $f_{P(13)}-{f_{D2}/2}\simeq 72.78 $~GHz and $f_{P(15)}- f_{D2}/2\simeq -88.91 $~GHz. Nevertheless, frequency transfer to the potassium lines is still attainable with the use of higher-order harmonics of a fiber phase modulator, as we shall see in the following section. For the potassium D1 line, the closest acetylene transition is the $\rm P(24)$, but its strength is significantly lower, and we could not detect it experimentally (see Fig.~\ref{Acetylene_spectrum}). In our case, we opted for the $\rm P(23)$ line ($1539.43$ ~nm), with a frequency difference $f_{P(23)}-{f_{D1}/2}\simeq 99.37 $ ~GHz.

\subsection{Experimental setup}\label{Sect2.2}

\begin{figure}[!h]
	\centering
	\includegraphics[width=0.9\linewidth]{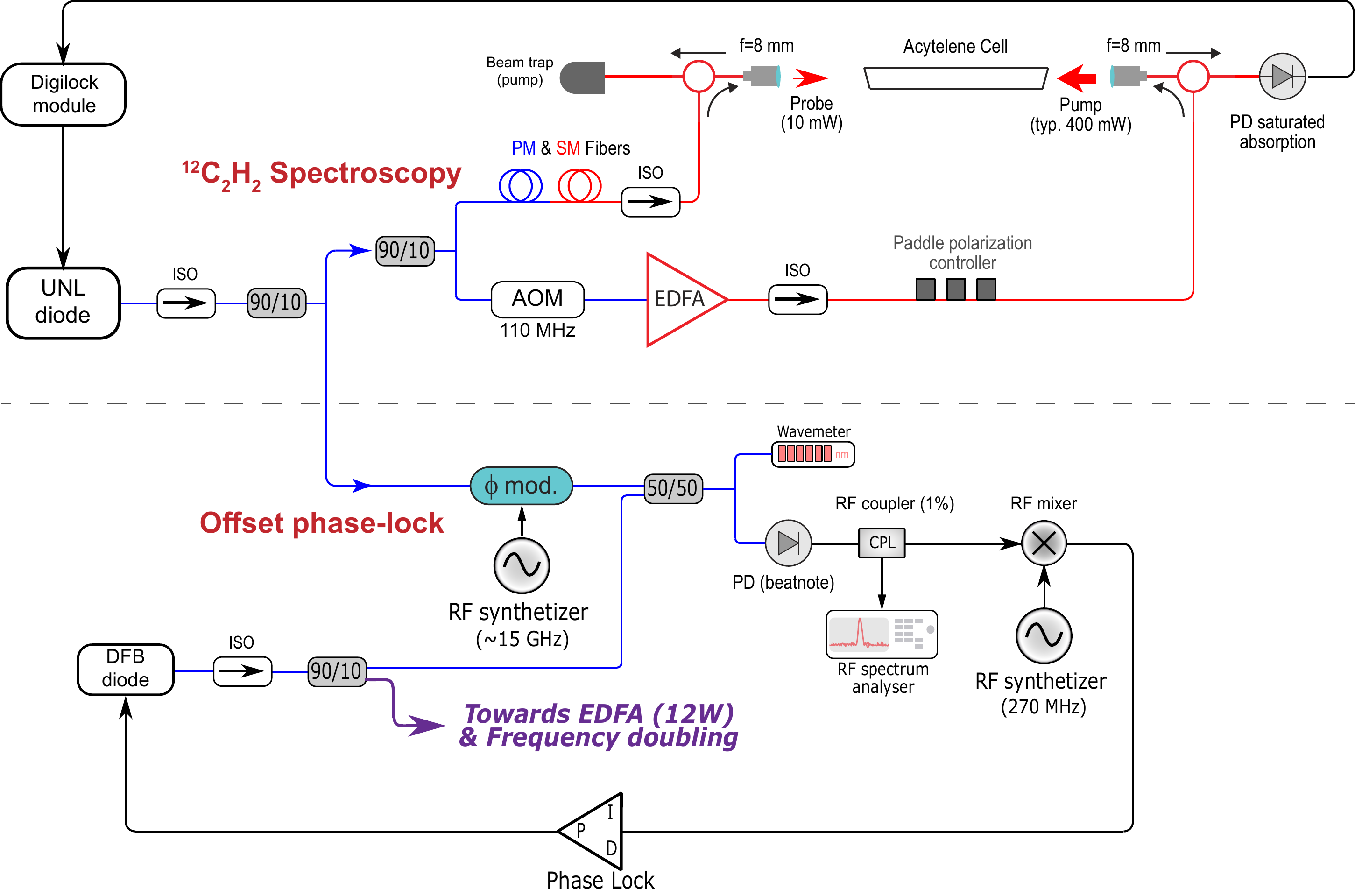}
	\caption{Overview of the experimental setup (AOM: Acousto-Optic Modulator; ISO: Fiber-Optic Isolator; SM (red): Single-Mode optical fiber; PM (blue): Polarization-Maintaining optical fiber; PD: Photo-diode; DFB: Distributed Feedback laser; EDFA: Erbium-Doped Fiber Amplifier). The system consists of two parts: the acetylene spectroscopy setup (upper half), for frequency stabilization of the UNL laser, and the offset phase-lock (lower half), which bridges the gap to twice the potassium cooling transition frequency, and seeds a high-power amplifier and SHG stage.}
	\label{setup}
\end{figure}

The experimental setup is presented in Fig.~\ref{setup}. It is composed of two main parts: the first part is used to stabilize the frequency of a ultra-narrow line (UNL) laser diode (`master') on a Doppler-free signal using a acetylene molecular transition in a low-pressure spectroscopy cell. The second part (described in Section~\ref{Sect2.4}) represents an offset phase-lock scheme, using a distributed feedback (DFB) laser diode (`slave'), which is used to `bridge' the $\sim 80$~GHz gap between the acetylene and the potassium atomic cooling transition(s), before SHG. In this way, we obtain a simple and compact, mostly fiber-based setup -- almost insensitive to vibrations and misalignments.

Our acetylene-based locking setup utilizes a commercial UNL laser diode (OE4023 model from OEwaves, using a whispering gallery mode resonator). The linewidth is $<1$~kHz, and it delivers $20$~mW through a polarization-maintaining (PM) optical fiber. This diode is temperature-tunable over $\sim 1$~nm, and has a fast current modulation ($100$~kHz bandwidth) input with a $\sim 200$~MHz range. A fiber isolator, protecting the diode against back-reflections, is followed by a PM fiber coupler which brings $10\%$ of the power towards the phase-lock setup. The $90\%$ remaining optical power is further separated, via a second PM coupler, to generate the pump and probe beams for saturated absorption. The probe beam ($90\%$ output, ~10 mW) passes through an optical  circulator and is directed into a low-pressure acetylene spectroscopy cell. The pump beam ($10\%$ output) passes through an Acousto-optic modulator (AOM) and is then amplified with a single-mode (SM) Erbium-Doped Fiber Amplifier (EDFA, IPG model EAD-500), generating a power up to $600$~mW. Additional SM fiber isolators are placed to further reduce optical feedback, for both the pump and the probe beams.

In the spectroscopy setup, we utilize a 50-cm-long commercial acetylene cell (Precision Glass Blowing, Colorado, USA) with wedged AR coated windows. The acetylene pressure is $50$~mTorr (manufacturer specification). Both the probe and the pump are collimated to $1/e^2$ diameters of $1.6$~mm, using $f=8.18$~mm aspheric lenses. They are sent in opposite directions through the cell, and then re-injected, with $90\%$ efficiency, in the opposite optical fibers. Using fibered optical circulators, the probe beam is separated after the cell and directed to a low-noise photodiode for detection, whereas the pump power is sent to a beam trap for termination. The acetylene cell and the two collimating lenses represent the only free-space part of the setup, and are all ruggedly mounted on a rigid cage system; no optical realignment was found necessary, even after several months of operation.

\begin{figure}[!h]
	\centering
	\includegraphics[width=0.9\linewidth]{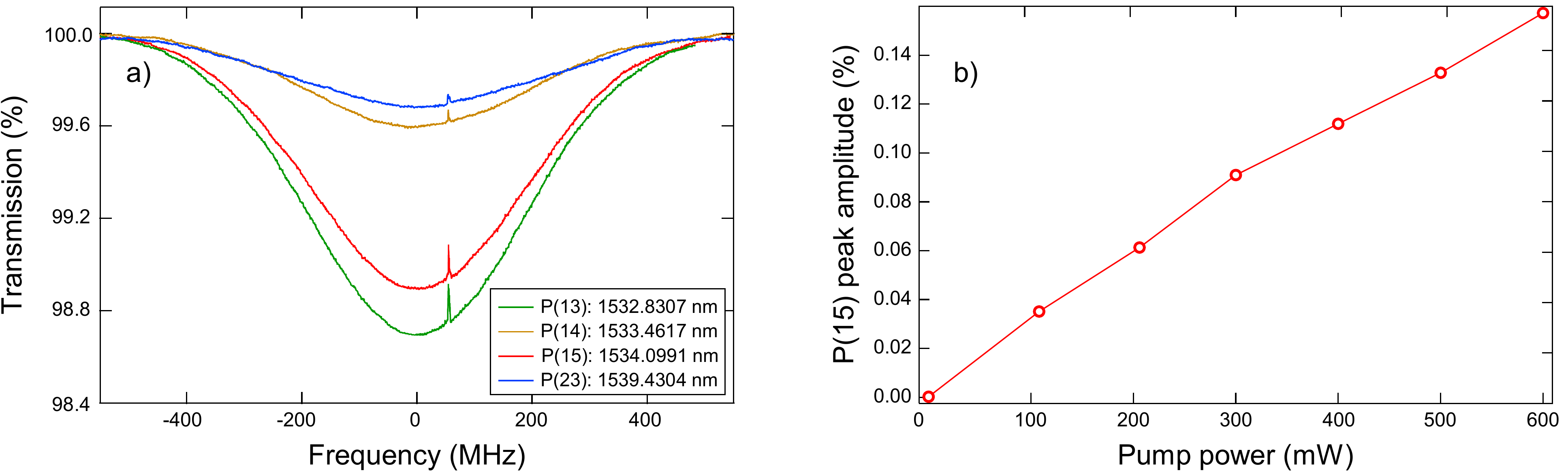}
	\caption{a) Normalized absorption spectra of four ro-vibrational transitions of the $^{12}\rm C_{2}H_{2}$ molecule.  We observe two strong Doppler lines, with $\sim 1\%$ absorption (P(13) and  P(15)) and two weaker lines with $\sim 0.3\%$ absorption (P(14) and  P(23)). In presence of the pump laser (400~mW power), the sub-Doppler saturated absorption peaks are observed at $f_{\rm AOM}/2=55$~MHz. b) Amplitude of the P(15) saturated absorption peak as a function of the power of the pump beam.}
	\label{fig:RaiesDoppler}
\end{figure}

We paid particular attention to avoid optical interferences on the spectroscopy detection scheme. The fiber-based part of the system has been therefore assembled by fusion splicing. However, residual internal reflections coming from the different fiber components have been found to alter the spectroscopy signal, on a level comparable to the absorption signal (i.e. with a contrast on the $1\%$ level). Two main improvements were made in order to address this issue. First, we found that using SM fiber components, whenever possible, we were able to decrease the amount stray internal reflexions. Second, we used an AOM to shift the frequency of the pump with respect to the probe by $f_{\rm AOM} = 110$~MHz. The fringes coming from crosstalk between the probe and the high-power pump will thus `self-average' on the detection photodiode, which allows us to directly observe the saturated absorption signal, in `single-shot' measurements.

\subsection{Acetylene saturated absorption spectroscopy and laser locking}\label{Sect2.3}

\begin{figure}[!h]
	\centering
	\includegraphics[width=0.8\linewidth]{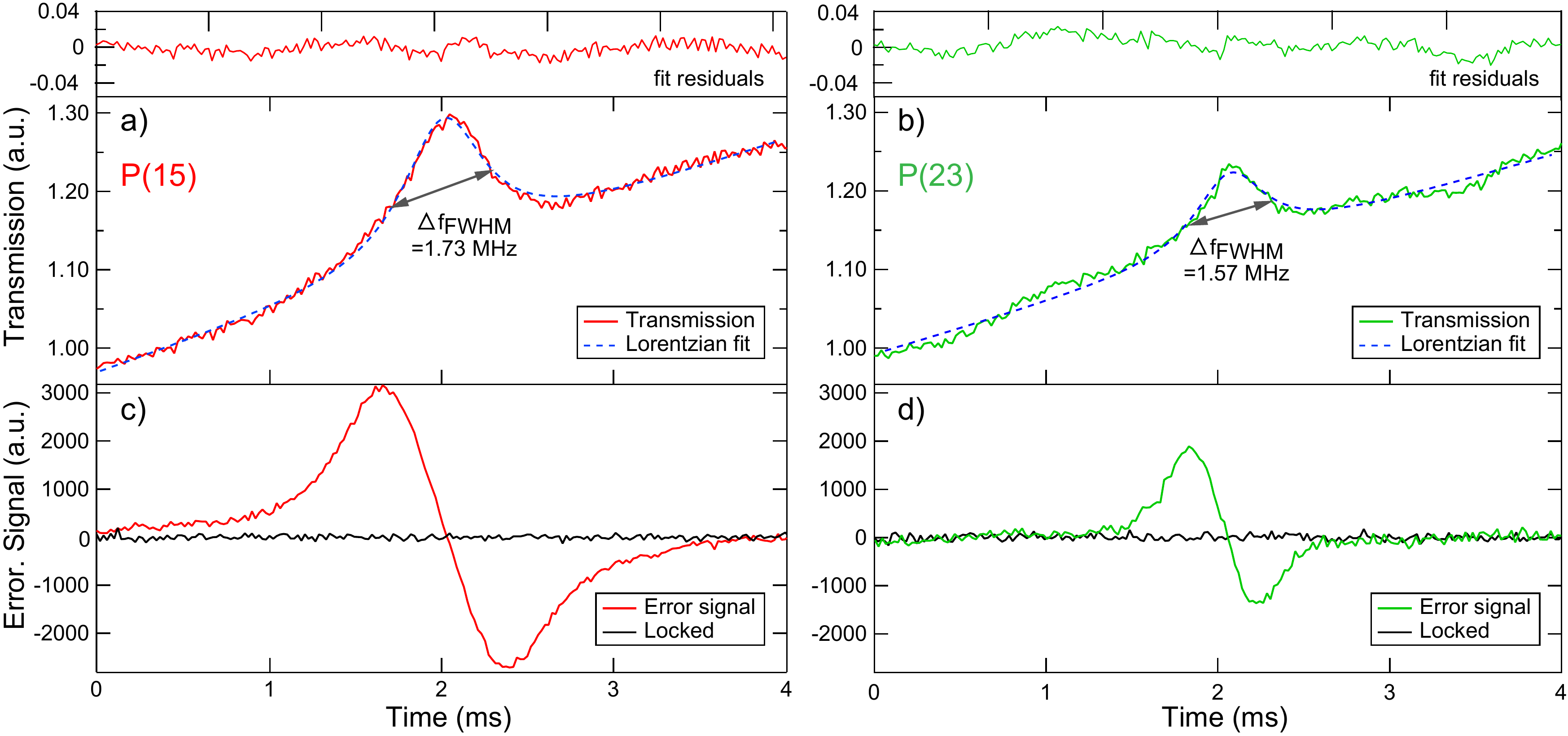}
	\caption{a) and b): Saturated absorption peaks of the P(15) and P(23) transitions, obtained by applying a linear scan to the UNL laser frequency  (rate 3 MHz/ms). The profiles are well fitted by a Lorentzian line shape (dashed blue lines: fit results; residuals in the top panel), and correspond to a FWHM of $1.73$~MHz for the P(15) transition and $1.57$~MHz in the case of the P(23) transition. A larger pump power of $600$~mW was used for the weaker P(23) transition, compared to $400$~mW in the case of the P(15) one. The saturated absorption peaks are used to generate the corresponding error signals, which are shown in c) and d). These signals are used to feed the PID filters, that generate the correction signal sent to the DFB diode laser. In black, we show the error signals of the laser, when locked to the corresponding acetylene peaks.} \label{fig:fig3}
\end{figure}

\begin{figure}[!h]
	\centering
	\includegraphics[width=0.6\linewidth]{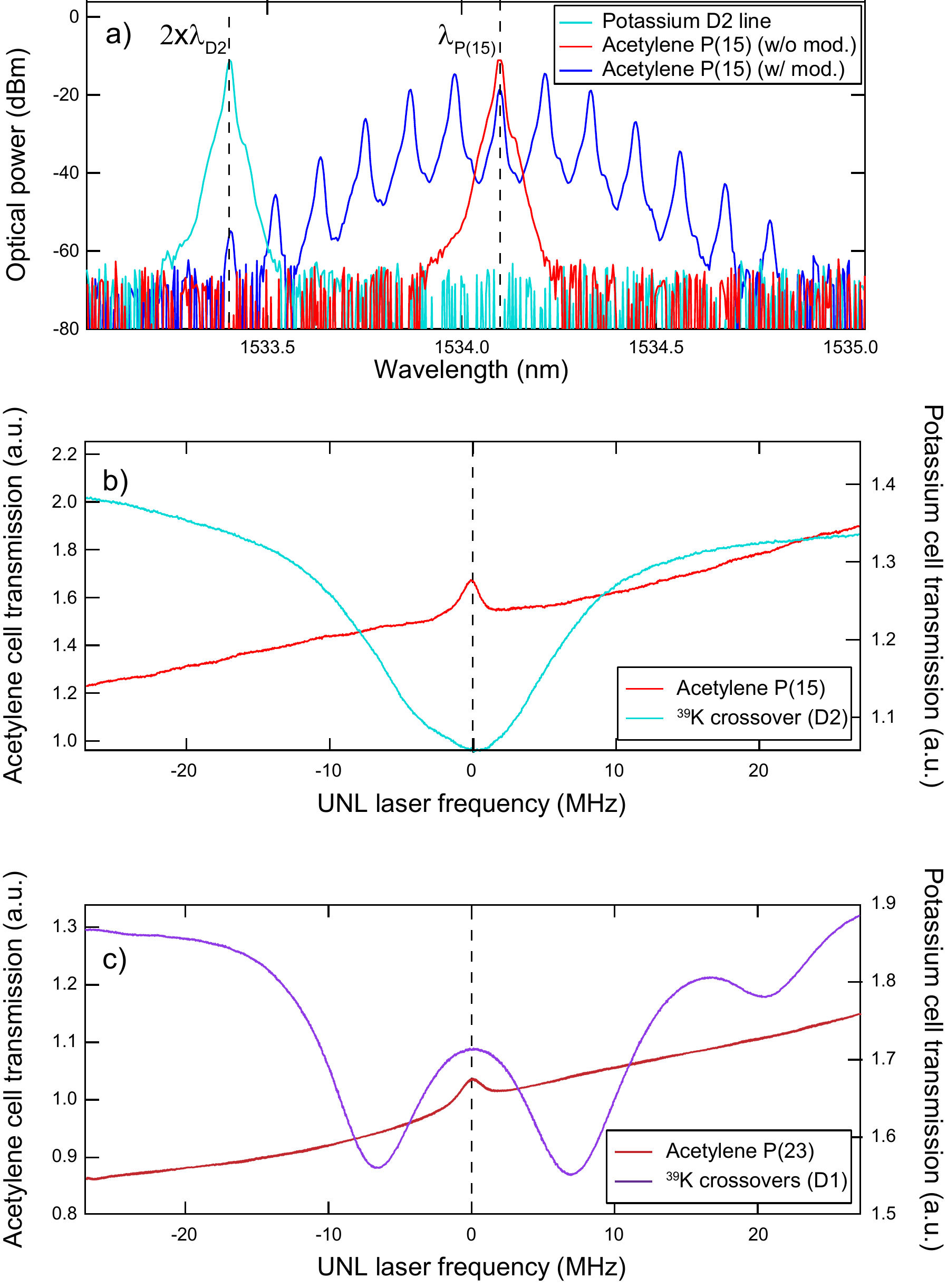}
	\caption{a): Optical spectrum of the UNL laser, before (red) and after (blue) the phase modulator. The modulation frequency is chosen such that the sixth lower sideband coincides with the DFB laser frequency (cyan), tuned to $2 \times \lambda_{\rm D2}$ transition. This allows to detect the low-frequency optical beat note between the two lasers, used for phase-locking. b): By scanning the UNL diode current, we superposed the acetylene saturated absorption peak (red) and the $^{39}$K D2 crossover dip (cyan), when the UNL and the DFB lasers are phase-locked. This corresponds to a phase modulator frequency of $15.119$~GHz. c): Same spectra obtained with a similar setup, using a 1540~nm UNL laser, showing the acetylene P(23) transition (purple) and the potassium D1 crossovers (magenta), which are well resolved, due to the larger hyperfine splitting of the $^2$P$_{1/2}$ state.}
	\label{fig:fig5}
\end{figure}

To characterize the spectroscopy setup, we temporarily replaced the UNL diode with a DFB laser, which provides a wider frequency scan range (via the laser current). This allowed us to observe different spectroscopic lines of interest for the potassium D2 line, shown in Fig.~\ref{fig:RaiesDoppler}(a). The Doppler-broadened P(15) line has a width of $440$~MHz (FWHM) and an amplitude of $1.1\%$. In presence of the pump beam, we observe a saturated absorption peak. Due to the relative frequency difference between the pump and the probe beams, its position is shifted by $f_{\rm AOM}/2=55$~MHz with respect to the center of the Doppler profile. Figure~\ref{fig:RaiesDoppler}(b) shows the amplitude of the saturated absorption peak as a function of the pump power. A typical power of 400 mW was chosen for regular operation, to preserve the lifetime of the pump amplifier, which yields a peak amplitude of $0.112\%$. Additional saturated absorption signals were observed, for the neighboring P(16) and P(17) lines. Using a different DFB laser, centered at 1540 nm, we also observed the $\rm P(23)$ line, with a corresponding $0.032\%$ amplitude of the saturated absorption peak.

Figure~\ref{fig:fig3} shows a zoom of the saturated absorption peaks of the $\rm P(15)$ line, obtained with the $1534$~nm UNL laser. This peak is well fitted with a Lorentzian shape (plus a quadratic background to take into account the Doppler profile at the position of the saturated absorption peak), with a FWHM linewidth of $1.73$~MHz. To generate the error signal for frequency locking of the UNL laser, we implement a modulation transfer scheme. For that, a sinusoidal frequency modulation at $125$~kHz is added to the $110$~MHz signal which drives the AOM of the acetylene saturated absorption pump beam. This modulation is detected by the probe beam photodiode and demodulated, using a digital laser lock-in module (Toptica DigiLock 110), which generates the error signal shown in Fig.~\ref{fig:fig3}. This error signal is fed back to the current of the UNL laser, which allows us to lock it to the acetylene saturated absorption peak. From the RMS amplitude of the locked signal, we infer an upper bound of $<50$~kHz for the linewidth of the laser locked to the acetylene peak. Using the same setup with a different UNL laser at $1540$~nm, we were also able to lock it to the $\rm P(23)$ line.


\subsection{Frequency transfer to NIR}\label{Sect2.4}

As discussed in section 2.1, the frequency difference between the P(15) line and $2 \times \lambda_{\rm D2}$ transition of potassium is $\sim88$~GHz. To bridge this gap, a phase modulator (Photline, model MPZ-LN-10) is implemented after the UNL laser which generate sidebands in the frequency spectrum, up to $\sim 100$~GHz (blue peaks in Fig.~\ref{fig:fig5}(a)). Using an AC-coupled photodiode, we detect the low-frequency (270 MHz) beatnote between the DFB diode and the lower sixth harmonic of the UNL laser, created by the phase modulator. The photodiode output is demodulated with a RF synthesizer, which generates the phase-lock error signal. This signal is sent to a high-bandwidth commercial PID module (Toptica FALC 110), whose output is summed, via a fast bias-T circuit, to the DFB diode current, which phase-locks it to the acetylene-stabilized UNL laser.

After frequency-doubling the DFB laser, we used a separate saturated absorption potassium vapor cell to observe the potassium atomic transitions. This is done by scanning the current of the UNL laser when the two diodes are phase-locked. The offset between acetylene and different potassium transitions can thus be calibrated, by changing the frequency of the phase modulator. Figure~\ref{fig:fig5}(b) shows the superimposed the P(15) line of the $^{12} \rm C_{2}H_{2}$ molecule (red) and the crossover transition~\cite{MudarikwaJPhysB2012} of the $^{39}$K (cyan), which is the most naturally-abundant isotope of potassium ($93.258\%$). This corresponds to a phase modulator frequency of $15.119$~GHz ($\pm1$~MHz). Additionally, we were able to duplicate the setup with different UNL and DFB diodes, tuned respectively to the acetylene P(23) and potassium D1 transitions, as shown in Fig.~\ref{fig:fig5}(c). In this case, several resolved crossover peaks are observed~\cite{Gozzini2015}, because of larger hyperfine splitting ($\sim 55$~Mhz) of the D1 $^2$P$_{1/2}$ excited state of $^{39}$K.

\subsection{Magneto-optical trapping of $^{41}$K atoms}

Finally, we used our frequency-locking setup to implement magneto-optical trapping of the $^{41}$K isotope of potassium. Although less naturally-abundant ($6.730\%$) than the $^{39}$K isotope, $^{41}$K was chosen in our experiment because of its favorable low-temperature collisional properties, i.e. a large and positive s-wave scattering length~\cite{FalkePRA2008}. This eliminates, for instance, the need of using Feshbach resonances in order to achieve Bose-Einstein condensation of $^{41}$K~\cite{KishimotoPRA2009}.

The frequency-stabilized DFB light is used to seed a high-power telecom fiber amplifier (Quantel, model ELYSA-A-1533-10-P-SN-M-CC), delivering up to $12$~W infrared light. The output of the amplifier is focused down to $80$~$\mu$m ($1/e^2$ diameter) onto a free-space SHG nonlinear crystal of PPLN, in a single-pass configuration, which generates the MOT light close to the $\sim767$~nm laser cooling transition ($^2$S$_{1/2},$ F$=2 \to ^2$P$_{3/2},$ F$=3$). In contrast with other alkaline atoms (Cs, Rb,...), potassium has a small hyperfine splitting  in the excited state ($\sim 17$~MHz for $^{41}$K, comparable to the natural linewidth of $6$~MHz), which leads to high depumping rates. For this reason, potassium MOTs require a large amount of repumper light - typically as much as the cooling light~\cite{KishimotoPRA2009}. To reduce power losses in free-space, in our setup the cooling and repumper frequencies are both generated, using fibered AOMs, and mixed together before the high-power fiber amplifier. This way, the output of the SHG crystal readily contains all laser cooling frequencies in the same spatial mode (avoiding the use of free-space AOMs, which typically generate important power losses). Using this setup, we obtain a power up to $2.2$~W for laser cooling, with a very good beam quality, compared e.g. with solid-state TA amplifiers.

\begin{figure}[!h]
	\centering
	\includegraphics[width=0.5\linewidth]{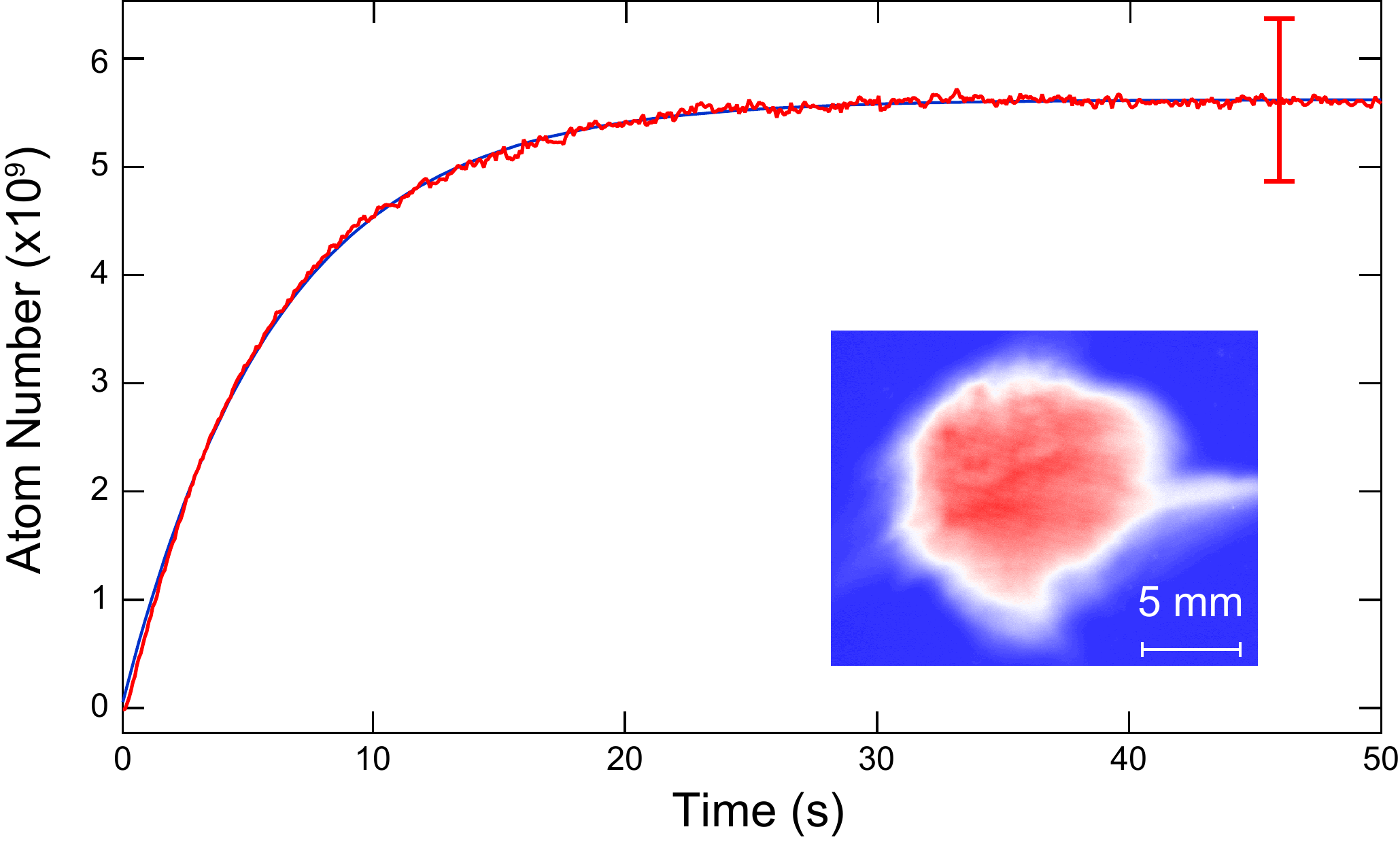}
	\caption{Loading evolution of the $^{41}$K atoms trapped in our 3D MOT. The blue curve is an exponential fit, which gives a $1/e$ loading time of $6$~s. Inset: fluorescence image of the $^{41}$K atom cloud.}
	\label{LoadingMOT}
\end{figure}

Our cold-atom system consists of a 3D MOT fed by a slow beam of atoms, produced using a commercial vapor-cell 2D MOT (developed by the LNE-SYRTE laboratory in Paris, France). The atomic source for the 2D MOT consists of a potassium ampoule connected to the 2D MOT chamber through a CF16 valve. A temperature gradient helps the potassium to migrate from the source (heated to 80 $^\circ$C) to the 2D MOT chamber (50 $^\circ$C). The vapor pressure in the chamber is maintained with a 2~l/s ion pump, at $\sim 5\times 10^{-8}$ mBar. The slow atomic beam is created by six pairs of retro-reflected elliptical beams (three for each transverse direction, with a total interaction length of $\sim10$~mm along the beam direction), together with a 2D magnetic field gradient of $12$~G/cm, produced by two pairs of coils. The 3D MOT is created in a second vacuum chamber, connected to the 2D MOT via a differential pumping stage, pumped by two getter-ion pumps (NexTorr, models D500 and D300). The pressure is on the order of a few $10^{-11}$~mBar, with a MOT lifetime $>$30 seconds. For the 3D MOT we use three pairs of independent laser beams, with $1/e^2$ diameters of $11.5$~mm and $\sim 40$~mW each. A relatively weak magnetic field gradient ($5$~G/cm along the axial direction of the coils) is used, to reduce light-assisted collisions processes, which limit the atoms number in potassium MOTs~\cite{PrevedelliPRA1999}. Using this configuration, we are able to trap up to $\sim 5.5 \times 10^9$ atoms of $^{41}$K. Figure~\ref{LoadingMOT} shows a loading curve of the MOT as function of the time, with an $1/e$ loading time of approximately $6$~s.

\section {Conclusion}
In conclusion, we demonstrated a versatile and robust fiber-based locking scheme of a telecom laser source directly applicable for potassium laser cooling. Our setup completely decouples the frequency locking from the amplification and SHG to NIR domain, which greatly widens the experimental possibilities. The laser system was validated by its successful implementation in a magneto-optical trapping setup for the $^{41}$K isotope, and should be easily extended, with minor changes, to the other isotopes of potassium. Further developments will include testing sub-Doppler cooling techniques for potassium using the D1 frequency locking setup on the P(23) acetylene transition. A particularly interesting improvement would consist in developing a fully fiber-based setup, by using acetylene-filled hollow-core optical fibers~\cite{BouyerHilico2019}. Their implementation would be straightforward in our setup. Thus, interesting perspectives could open towards mobile cold atom setups, with possible metrological applications using potassium isotopes~\cite{Antoni-MicollierPRA2017}. Finally, the offset phase-locking technique can be useful for developing acetylene-based frequency standards, spanning the entire $1515$ -- $1540$~nm range.

\section*{Funding}
This work was financially supported by Agence Nationale de la Recherche through Research Grants K-BEC No. ANR-13-BS04-0001-01 and MANYLOK No. ANR-18-CE30-0017, the Labex CEMPI (Grant No. ANR-11-LABX-0007-01), the I-SITE ULNE / ANR-16-IDEX-0004 ULNE, the Programme Investissements d'Avenir ANR-11-IDEX-0002-02, reference ANR-10-LABX-0037-NEXT, the Ministry of Higher Education and Research, Hauts-de-France Council and European Regional Development Fund (ERDF) through the Contrat de Projets Etat-Region (CPER Photonics for Society, P4S).

\section*{Acknowledgments}
The authors thank Leticia Tarruell for useful discussions about the $^{41}$K MOT setup.

\section*{Disclosures}
The authors declare no conflicts of interest.


\begin{thebibliography}{10}
\bibitem{Charbel2019InPrep}
	The complete details of this amplification stage is out of the scope of this
	paper, and will be provided in a future work.
\end{thebibliography}

\end{document}